\documentclass[pra,twocolumn,superscriptaddress,amsmath,amssymb,floatfix]{revtex4-2} 
\usepackage{graphicx} 
\usepackage{bm}       
\usepackage[colorlinks=true,linkcolor=blue,urlcolor=blue,citecolor=blue,linkcolor=blue,pdfstartview=FitH]{hyperref}

\usepackage{enumerate}
\usepackage{amsmath}
\usepackage{xcolor}
\usepackage{soul}
\usepackage{cases}
\usepackage{microtype}
\usepackage{siunitx}
\usepackage{subdepth}
\usepackage[utf8]{inputenc}
\usepackage[T1]{fontenc}
\usepackage{txfonts} 

\begin{document}

\title{Intermittent Motility of a Synthetic Active Particle in Changing Environments}

\author{Rudra Sekhri}
\affiliation{Optical Sciences Centre, Swinburne University of Technology, Melbourne 3122, Australia}
\author{Rahil N. Valani}
\affiliation{Rudolf Peierls Centre for Theoretical Physics, Parks Road,
University of Oxford, OX1 3PU, United Kingdom}
\author{Tapio Simula}
\affiliation{Optical Sciences Centre, Swinburne University of Technology, Melbourne 3122, Australia}

\begin{abstract}

We experimentally investigate the dynamics of synthetic active particles composed of gravitationally bouncing, superwalking droplets confined within an annular fluid bath. Driven by a topologically pumping dual-frequency waveform, the droplets exhibit alternating active (walking) and dormant (bouncing) phases, producing intermittent azimuthal motion. Tracking individual droplets reveals pseudolaminar chaotic dynamics in the time series of particle's angular position, characterized by laminar plateaus that are interrupted by short irregular bursts of activity. Increasing the driving amplitude induces a qualitative change in the active particle's intermittent dynamics, arising from a symmetry-breaking transition in its Faraday-wave field environment: continuous $\mathrm{SO}(2)$-symmetric “channelling” waves give way to discrete “trapping” patterns. These findings demonstrate how environmental symmetry and spatiotemporal structure modulate motility and intermittency in synthetic active matter.
\end{abstract}

\maketitle

\section{Introduction}

An active particle is a non-equilibrium entity that converts energy into directed motion. Examples span from living organisms—cells, bacteria, birds, humans—to synthetic systems such as electronic robots~\cite{Palagi2018} and chemically active colloids~\cite{doi:10.1146/annurev-conmatphys-031214-014710}. In many such systems, motility does not remain constant in time but undergoes intermittent or cyclical modulation where the active particles alternate between active and quiescent phases. For example in living systems, the 24-hour day–night circadian rhythm induces periodic changes in foraging behavior (diurnal versus nocturnal animals). Other examples include bacterial run-and-tumble dynamics~\cite{Berg1972}, and light-field-driven microswimmers that periodically halt and restart motion~\cite{Arlt2018,Frangipane2018}. In natural environments, even simple microorganisms can adapt their run speed or tumbling frequency in response to spatiotemporal variations in nutrient or chemical gradients, enabling efficient chemotactic navigation~\cite{Celani2010}. Theoretically, such behavior has been captured using stochastic switching models~\cite{Datta2024} and they exhibit rich transport behaviors, ranging from subdiffusive to superdiffusive regimes, depending on the statistics of switching times and environmental constraints. Because such intermittent and adaptive motility is ubiquitous across living active matter, it is natural to ask: \textit{how do synthetic active particles modulate their intermittent motility in response to environments that vary in both space and time?} In this work, we address this question using a synthetic active system of a superwalking droplet~\citep{Valani2019a}.

When a bath of silicone oil is vertically vibrated at a single frequency, a droplet of the same liquid can be made to periodically bounce and walk horizontally across the surface~\citep{Couder2005WalkingDroplets}. Such a walking droplet, or \textit{walker}, exists below the Faraday instability threshold~\citep{Faraday1831a}, where the bath surface remains globally flat except for the localized perturbations created by the repeated droplet impacts. Above this threshold, the surface itself destabilizes forming standing Faraday waves, and the droplet's resonant bouncing and walking motion gets strongly perturbed. Below the Faraday instability in the walking state, during each bounce, the walker excites a localized, slowly decaying standing wave that it interacts with on subsequent impacts, giving rise to a self-propelled, steady horizontal motion. The droplet and its accompanying wave field coexist as a \emph{wave–particle entity}: the droplet excites the waves, and the waves in turn guide the droplet’s motion. This mutual coupling yields a macroscopic realization of a system comprising both particle and wave characteristics. Interestingly, these walking droplets have been shown to mimic several peculiar features that are usually associated with the quantum world~\citep{Bush2020review,Bush2024}.

The walker is active in the sense of active matter~\citep{PismenActiveMatterBook}, as it locally harvests energy from the globally driven bath and converts it into directed motion. Thus, walking droplets constitute a unique class of synthetic, inertial, active particles propelled by self-generated fields. When the bath is driven simultaneously at two frequencies, $f$ and $f/2$, a new regime of larger and faster walkers emerges, known as \textit{superwalkers}~\citep{Valani2019a,superwalkernumerical}. Their steady propulsion speed can be tuned by varying the phase difference $\Delta\phi$ between the two driving tones. Furthermore, by imposing time-dependent phase modulations $\Delta\phi(t)$, one can realize topological pumping in the time domain~\cite{topopump}.

\begin{figure}
    \centering
    \includegraphics[width=\linewidth]{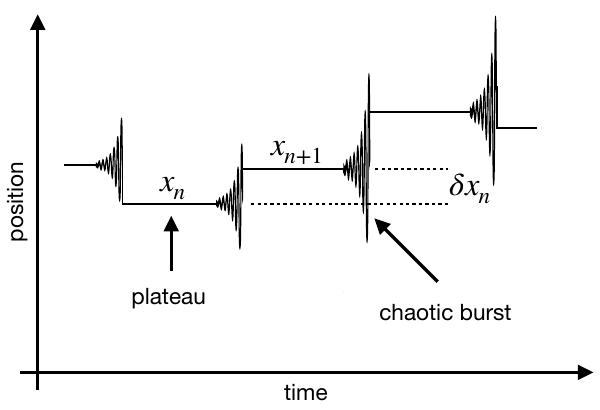}
    \caption{Illustration of intermittent motility of the active particle. Intermittent motility showing features of pseudolaminar chaos when the particle position is shown as a function of time. The time series comprises of consecutive low-motility, laminar plateaus $x_n$ and $x_{n+1}$ with $\delta x_{n}=x_{n+1}-x_{n}$, separated by high motility, irregular bursts.
    }
    \label{fig:PLC1}
\end{figure}

In experiments, superwalking droplets have been observed to exhibit a periodic modulation of their motility, termed the \textit{stop-and-go} motion~\citep{Valani2019a}. This behavior arises when the two driving frequencies are slightly detuned, i.e., $f$ and $f/2 \pm \epsilon$, where $\epsilon \ll f$. Such detuning has the same effect as a slowly varying phase difference $\Delta\phi(t)$ between the two driving tones $f$ and $f/2$, causing the system to periodically alternate between a stationary bouncing droplet state and a steady walking state~\citep{ValaniSGM}. The resulting long-timescale modulation of the driving thus produces a rhythmic sequence of “stop” and “go” phases in the droplet’s motion—an emergent form of periodic, intermittent motility in a synthetic active particle. Theoretical models of superwalking droplets under such detuned forcing predict that these periodic transitions in time can give rise to either spatially regular trajectories, such as back-and-forth or forth-and-forth motion, or spatially irregular motion~\citep{ValaniSGM}. When the spatial motion is irregular, the time series of the droplet position obtained from solving minimal models displays a type of intermittent motility identified as \textit{pseudolaminar chaos}~\citep{pseudolaminar2023,cwjk-n45m}. As illustrated schematically in Fig.~\ref{fig:PLC1}, pseudolaminar chaos is characterized by piecewise-constant time series, where nearly constant “laminar” phases corresponding to a stationary droplet are intermittently interrupted by “irregular” bursts of walking motion.

In this work, we experimentally investigate this intermittent motility associated with pseudolaminar chaos in topologically pumped stop-and-go superwalking droplets. We focus on the regime of high driving amplitudes, where the forcing periodically drives the system across the Faraday instability threshold~\citep{Faraday1831a,benjamin1954}, leading to the periodic growth and decay of standing Faraday waves on the liquid surface. Since the pattern selection of Faraday waves can vary with system parameters, we obtain different types of intermittent motility that are dictated by the symmetry of the underlying Faraday waves, which we explore here.

The remainder of this paper is organized as follows. Section~\ref{sec:methods} describes the experimental setup and data analysis procedures. Section~\ref{sec:results} presents the main results on intermittent motility and the symmetry-breaking transition of the Faraday wave field. Finally, Sec.~\ref{sec:conclusions} summarizes the findings.

\section{Methods}
\label{sec:methods}

\begin{figure}[!t]
    \centering
    \includegraphics[width=\linewidth]{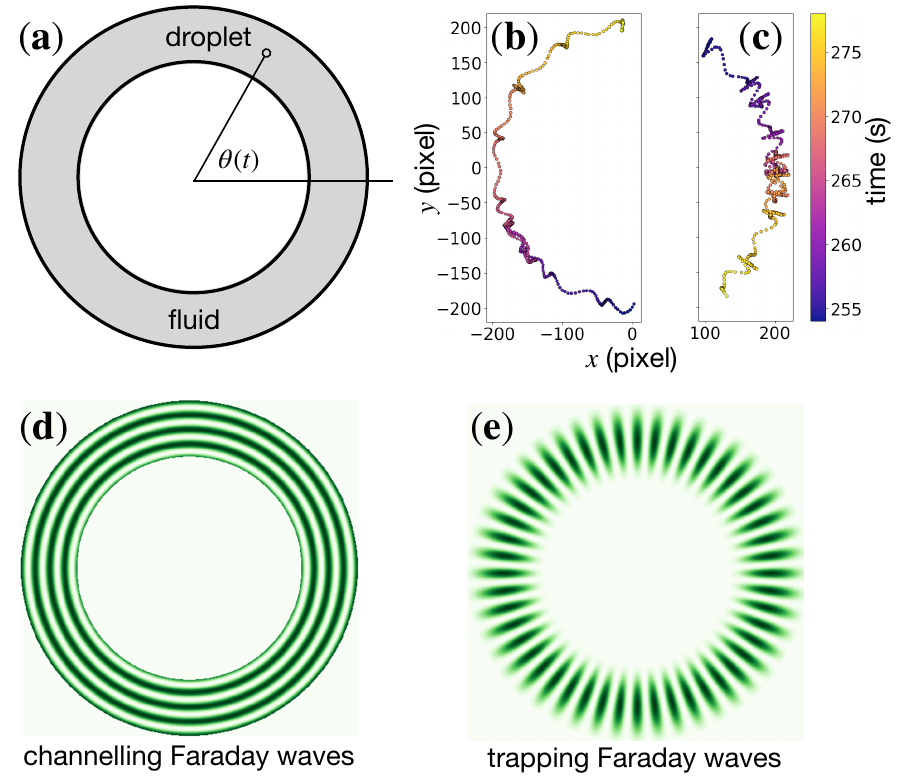}
    \caption{Geometric characteristics of the experiments. The droplet is confined within an annular region (a) of fluid bath with inner radius of 25.5 mm, channel width of 12.5 mm, and 2-4 mm depth of siloxane with 20 cSt kinematic viscosity. A typical droplet trajectory shown in (b) for $a_{\rm rms} =2.18g$ consists of small radial vibrations and azimuthal speed variations. By contrast, for $a_{\rm rms} =2.48g$ (c) strong radial vibrations take place with shorter strides in between. The 40Hz and 80Hz sinusoidal drive components, respectively, have independent Faraday thresholds and induce (d) radial (channelling) and (e) azimuthal (trapping) Faraday wave patterns when the respective thresholds are exceeded. The environments (d) and (e) lead to behaviours observed in (b) and (c), respectively.}
    \label{fig:PLC2}
\end{figure}

The general characterization of our apparatus is provided in \cite{SUTexpt}. Specific to the experiments in this work, superwalking droplets of radius $R_{\rm drop}\approx 0.8$ mm were deposited on the fluid bath. The topography of the fluid bath was modified in such a way that the droplets were trapped within a quasi-1D annular region, see Fig.~\ref{fig:PLC2}(a). The annulus has an inner radius of 25.5 mm, width of 12.5 mm and depth of 4.5 mm. This channel is filled with siloxane of 20 cSt kinematic viscosity to a depth of 2-4 mm such that a meniscus that forms in the channel creates a curved surface that traps the droplets in the radial direction. The annular bath realizes periodic boundary conditions along the azimuthal direction simulating an infinite one-dimensional system, thereby allowing extended observation periods of the superwalkers as the droplet moves along the annular channel. 

The fluid bath is driven sinusoidally along the axis of gravity by a waveform 
\begin{equation}
a_{\rm drive}(t) = a_{80}\sin(2\pi f_{80}t)+a_{40}\sin(2\pi f_{40}t + \Delta \phi(t))
\label{eq:drive}
\end{equation}
with frequencies $f_{80} = 80\,\mathrm{Hz}$ and $f_{40} = 40\,\mathrm{Hz}$ and an amplitude ratio $a_{40}/a_{80}=5/14$ and phase difference $\Delta \phi(t)=\pi t/2$ rad. Our key control parameter is the measured root-mean-square (RMS) acceleration of the bath $a_{\rm rms} = \sqrt{\int_{t}^{t+\Delta t} |a_{\rm mea}(t)|^2 dt /\Delta t}$, where the averaging is done over $\Delta t=1s$  window and $a_{\rm mea}(t)$ is an instantaneous value obtained from an accelerometer mounted on the fluid bath. The value of $a_{\rm rms}$ is adjusted by a linear amplifier that drives the electrodynamic shaker which is mechanically connected to the fluid bath \cite{SUTexpt}.

We use top view imaging to track the position and velocity of the droplets. Figures \ref{fig:PLC2}(b) and (c) show two samples of the detected raw $(x,y)$ pixel positions of a droplet. By varying the overall RMS amplitude $a_{\rm rms}$ of the periodic driving waveform, different behaviours of the droplet are found to emerge. The key derived observable in this work is the angular position $\theta(t)$ of the droplet as a function of time. This angular position is extracted from the continuously streamed video frames recorded at $30$ fps and processed using OpenCV's particle tracking library \cite{opencv}. This allows us to retain only the raw Cartesian $(x,y)$ pixel position of the droplet to reduce the data storage requirements. The instantaneous angular position $\theta = \arctan(y/x)$ of the droplet is derived from the detected Cartesian coordinates. 

\section{Results}
\label{sec:results}

When the bath is driven by a waveform defined by Eq.~(\ref{eq:drive}) at an amplitude between the superwalking threshold and the Faraday instability thresholds, a freely superwalking droplet exhibits a characteristic \textit{stop-and-go} motion on the surface of the fluid. In this regime, the droplet periodically crosses the transition between steady walking and stationary bouncing~\cite{Valani2019a,ValaniSGM}. During the “go” phase, the droplet translates horizontally across the surface, while during the “stop” phase it bounces in place without net motion, representing alternating phases of activity and dormancy in a synthetic active system.

Previous studies have shown that when a droplet is fully trapped within a small circular region and the analysis is restricted to its vertical dynamics, the system can exhibit topological pumping in the time domain~\cite{topopump}. In contrast, in the present experiments the droplet is confined only in the radial direction, allowing it to move freely along the azimuthal coordinate. The root-mean-square driving amplitude is tuned such that Faraday waves periodically emerge in the annular channel [see Figs.~\ref{fig:PLC2}(d) and \ref{fig:PLC2}(e)]. This spatiotemporal forcing modifies the droplet’s dynamics, producing a coupled motion characterized by periodic stop-and-go behavior along the azimuthal direction, accompanied by vibrational motion in the radial direction [Figs.~\ref{fig:PLC2}(b) and (c)]. In our system, the plateaus in Fig.~\ref{fig:PLC1} correspond to the radial jittering of the droplet position in Figs.~\ref{fig:PLC2}(b) and (c) and the chaotic bursts correspond to the smoother wiggly trajectories where the droplet moves fast during the go phase.

\begin{figure}[!t]
    \centering
    \includegraphics[width=\linewidth]{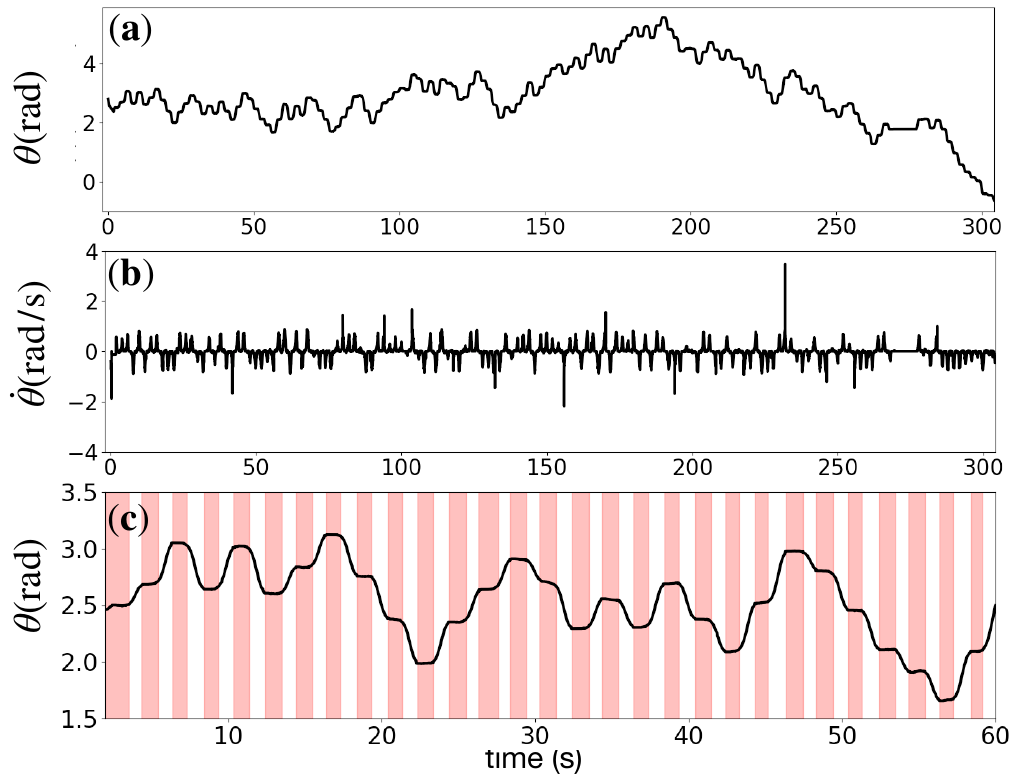}
    \caption{Level detection from the recorded droplet time series showing (a) the angular position $\theta(t)$ and (b) the angular speed $\dot{\theta}(t)$ of a droplet within the circular channel as functions of time. (c) angular position in a zoomed in region of panel (a) where the pink shaded regions indicate the detected time intervals during which the angular position of the droplet is not changing, corresponding to $\dot{\theta}(t)=0$ and constant $\theta(t)$. 
    }
    \label{fig:PLC3}
\end{figure}

Figure~\ref{fig:PLC3}(a) and (b) show, respectively, the angular position and angular velocity of a droplet for $a_{\rm rms}=2.18g$. 
To identify the intermittent stop-and-go dynamics, the $\theta(t)$ signal is first smoothed using a one-dimensional Gaussian filter to suppress high-frequency imaging noise, and then processed with a threshold-based filter to detect stationary plateaus corresponding to the “stop” phases of motion. These plateaus define discrete angular levels $\theta_n$, where $n$ denotes the integer level index. The detected stationary levels are highlighted in pink in Fig.~\ref{fig:PLC3}(c) where a zoomed-in view of $\theta(t)$ is shown.

The time series exhibits a clear alternation between extended laminar (stationary) intervals and short irregular bursts of motion, consistent with the intermittent dynamics characteristic of pseudolaminar chaos~\citep{pseudolaminar2023}. While such features are qualitatively apparent in the angular trajectories, the influence of the spatially structured environment on motility can be quantitatively characterized by analyzing the correlations of the level separations $\delta\theta_n = \theta_{n+1} - \theta_n$, as discussed below.

\begin{figure*}
    \centering
    \includegraphics[width=\linewidth]{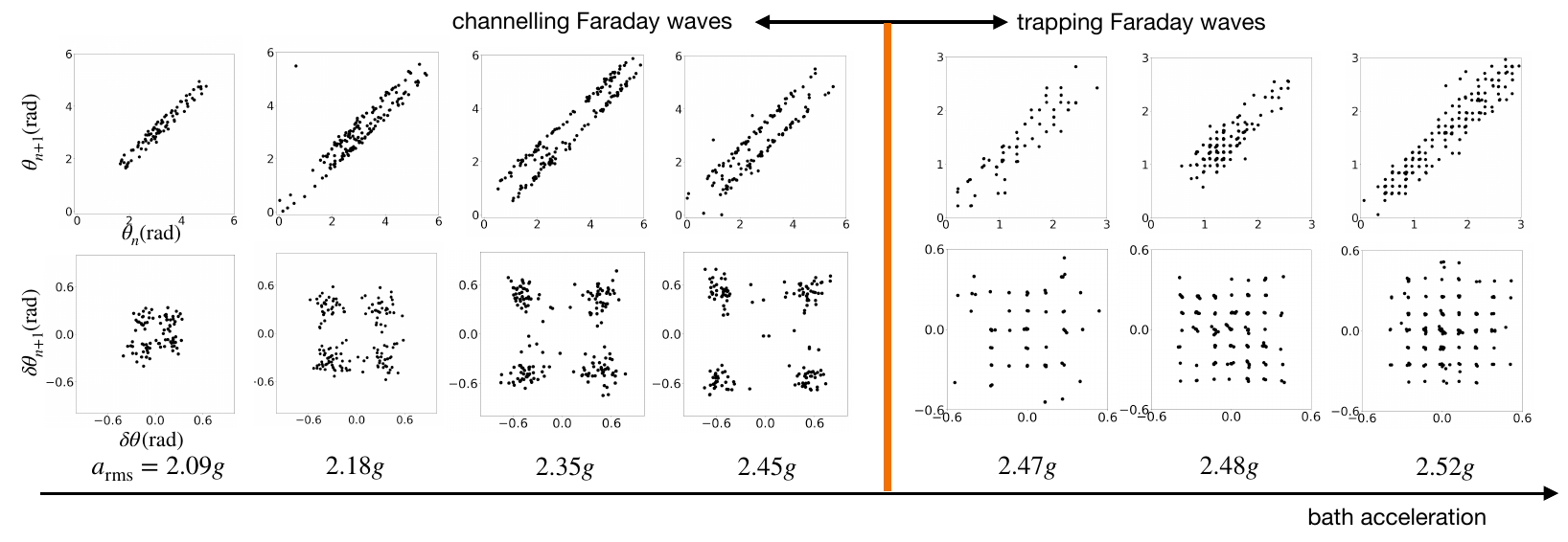}
    \caption{The first return maps of $\theta$ and $\dot{\theta}$ for increasing bath acceleration. The vertical bar indicates the threshold below which intermittent channelling Faraday waves, see Fig.~\ref{fig:PLC2}(c), emerge in the annular fluid bath. Above this threshold, the intermittent trapping Faraday waves are observed, Fig.~\ref{fig:PLC2}(d). The angular scales are different for different values of $a_{\rm rms}$ because the free walking speed of the droplet increases with increasing $a_{\rm rms}$. 
    }
    \label{fig:PLC6}
\end{figure*}

Once the discrete angular levels, $\theta_n$, have been identified, we construct the corresponding first-return maps for both the angular position and the level separation. Figure~\ref{fig:PLC6} presents the first-return maps of the angular position (top row) and the level separation $\delta\theta_n$ (bottom row) as functions of the root-mean-square bath acceleration $a_{\rm rms}$. We consider in detail the column corresponding to $a_{rms}=2.35g$, where the top panel shows the angular position at a given level, $\theta_{n+1}$, plotted as a function of the position at the preceding level, $\theta_n$, and the bottom panel shows an analogous return map for the level separation $\delta\theta_n = \theta_{n+1} - \theta_n$. The position return map exhibits two prominent, approximately linear correlation stripes separated by about one radian and symmetrically arranged about the diagonal. These two branches correspond to the droplet’s tendency to move alternately in the clockwise or counterclockwise direction following each dormant phase. The return map of the level separations displays a corresponding fourfold symmetric structure, consistent with this bidirectional switching behavior. The position return map in the top plot closely resembles those reported in a one-dimensional Lorenz-like model for stop-and-go superwalkers (see Fig.~2 of ~\cite{pseudolaminar2023}). This alternation between clockwise and counterclockwise motion is reminiscent of a one-dimensional random walk with a fixed step length, where each laminar phase defines a discrete step and the signed displacement $\delta\theta_n$ represents the step size. The distribution of $\delta\theta_n$ is approximately symmetric about zero, reflecting the absence of a net directional bias in the long-time average. 

For driving amplitudes below the threshold value $a_{\rm rms} \approx 2.46g$, the return maps show similar qualitative features,
characterized by two well-defined correlation stripes indicative of bidirectional switching between clockwise and counterclockwise motion. The expansion of the observed structures in the return maps is a direct consequence of the free-walking speed of the droplet increasing with increasing $a_{\rm rms}$ \cite{Valani2019a} leading to the greater average angular distances between consequtive plateaus. As the driving amplitude is increased beyond a threshold, the structure of the return maps changes markedly. The simple two-stripe pattern gives way to a fine-grained, multi-valued grid-like organization. 

The observed transition in the droplet dynamics can be directly linked to changes in the underlying Faraday wave field sustained by the system. For driving amplitudes below the threshold, $a_{\rm rms} \lesssim 2.46g$, the bath periodically supports circular standing Faraday waves primarily excited by the 80~Hz component of the drive [see Figs.~\ref{fig:PLC2} (b) and (d)]. We refer to these as \textit{channelling} Faraday waves, as their continuous $\mathrm{SO}(2)$ rotational symmetry effectively preserves isotropy along the azimuthal direction. As a result, they exert only a weak modulation on the droplet’s horizontal motion, allowing the droplet to travel a nearly constant angular distance of approximately $0.5$~rad between consecutive stop phases. In this regime, the droplet dynamics closely resemble those of the freely walking stop-and-go motion, despite the presence of periodically emerging Faraday waves. During the stop phase, the droplet is gently rocked radially by the Faraday waves.  

When the driving amplitude exceeds the threshold, $a_{\rm rms} \gtrsim 2.46g$, the 40~Hz component of the drive dominates, giving rise to a new class of Faraday waves whose wavelength is roughly twice that of the channelling waves. Importantly, these \textit{trapping} Faraday waves break the continuous $\mathrm{SO}(2)$ symmetry, leading to a discrete azimuthal wave pattern that forms a series of 48 localized minima or dynamical “traps” along the annular channel [see Figs.~\ref{fig:PLC2}(c) and (e)]. The droplet can become transiently confined within any of these dynamically generated traps when Faraday waves are present in the system, resulting in the spontaneous emergence of intermittent localization along the azimuthal direction. When trapped, the droplet undergoes rapid jittering motion in the radial direction [see Figs.~\ref{fig:PLC2}(c)]. The azimuthal droplet dynamics thus mimic a one-dimensional discrete random walk, with each trap functioning as a lattice site and the droplet stochastically hopping between them with a variable step length.

This transition in the spatial environment from channelling to trapping waves thus represents a symmetry-breaking bifurcation in the spatiotemporal environment, which in turn drives a qualitative change in the motility landscape experienced by the active droplet.

\section{Conclusions}
\label{sec:conclusions}

In summary, we have investigated the dynamics of periodically driven synthetic active particles composed of superwalking siloxane droplets confined within an annular fluid bath. These droplets exhibit alternating active (walking) and dormant (bouncing) phases, giving rise to a characteristic stop-and-go motility. By analyzing the level sets derived from the azimuthal position time series, we constructed first-return maps of the angular position and level separation, which display features closely resembling those known to characterize pseudolaminar chaos. Upon increasing the driving amplitude, we observed a structural transition in the droplet's intermittent motility associated with an $\mathrm{SO}(2)$ symmetry breaking of the underlying Faraday-wave field.  

Our results demonstrate how symmetry and spatiotemporal structure in the dynamical environment can profoundly influence the behaviour of synthetic active matter. The transition from continuous to discrete symmetry in the Faraday landscape effectively transforms the droplet motion from continuous to lattice-like random-walk dynamics. More broadly, these findings highlight the role of environmental symmetry breaking as a mechanism for generating intermittent and correlated motility in active systems, offering new routes to control active particle transport through engineered spatiotemporal potentials.

\begin{acknowledgements}
We acknowledge support from the Leverhulme Trust [Grant No. LIP-2020-014] (R.V.) and the Australian Research Council (ARC) Future Fellowship [Grant No. FT180100020] (T.S.). 
\end{acknowledgements}

\bibliography{bibliography}
\bibliographystyle{apsrev4-2}

\end{document}